\documentclass[prl,twocolumn,floatfix,showpacs,preprintnumbers,superscriptaddress,amsmath,amssymb]{revtex4}
\usepackage{graphicx}

\begin{document}

\title{Electrical transport through single-molecule junctions: from molecular orbitals
to conduction channels}

\author {J. Heurich} 
\affiliation{Institut f\"ur theoretische Festk\"orperphysik, Universit\"at Karlsruhe, 
76128 Karlsruhe, Germany}

\author {J.C. Cuevas}
\affiliation{Institut f\"ur theoretische Festk\"orperphysik, Universit\"at Karlsruhe, 
76128 Karlsruhe, Germany}

\author {W. Wenzel} 
\affiliation{Forschungszentrum Karlsruhe, Institut f\"ur Nanotechnologie,
76021 Karlsruhe, Germany}

\author {G. Sch\"on}
\affiliation{Institut f\"ur theoretische Festk\"orperphysik, Universit\"at Karlsruhe, 
76128 Karlsruhe, Germany}
\affiliation{Forschungszentrum Karlsruhe, Institut f\"ur Nanotechnologie,
76021 Karlsruhe, Germany}

\date{\today}

\begin{abstract}
We present an atomistic theory of electronic transport through single
organic molecules that reproduces the important features of the
current-voltage (I-V) characteristics observed in recent experiments. We
trace these features to their origin in the electronic structure of
the molecules and their local atomic environment. We demonstrate
how conduction channels arise from the molecular orbitals and elucidate
which specific properties of the individual orbitals determine their
contribution to the current.
\end{abstract}


\maketitle

{\em Introduction.}--- Present trends in the miniaturization of
electronic devices suggest that ultimately single molecules may be
used as electronically active elements in a variety of
applications~\cite{Petty95,Joachim00}. Recent advances in the
manipulation of single molecules now permit to contact an individual
molecule between two electrodes (see Fig.~1) and measure its
electronic transport
properties~\cite{Joachim95,Datta97,Reed97,Kergueris99,Dekker00,Reichert01}.
In contrast to single-electron transistors based on metallic islands
\cite{Grabert92}, molecular devices have a more complicated, but in
principle tunable, electronic structure.  Interesting and novel
effects, such as negative differential conductance~\cite{Chen99}, were
observed in some of these experiments, which still, by-and-large, beg
theoretical explanation. In addition to generic principles of nanoscale
physics, e.g. Coulomb blockade~\cite{Kergueris99,Banin99,Park00},
the chemistry and geometry of the molecular junction emerge as the 
fundamental tunable characteristics of molecular
junctions~\cite{Joachim95,Datta97,Emberly98,Yaliraki99,Ventra00,Taylor01,Palacios01}.

In this paper we present an atomistic theory that bridges traditional
concepts of mesoscopic and molecular physics to describe
transport through single organic molecules in qualitative agreement
with recent break-junction experiments~\cite{Reichert01}. We combine
\emph{ab initio} quantum chemistry calculations with non-equilibrium Green 
functions techniques to illustrate the emergence of {\em conduction channels} 
in a single-molecule junction from the molecular
orbitals (MO). We further show how the specific properties of individual MOs
are reflected in their contribution to the current. Using this data
we provide insight into the microscopic origin of the nonlinear I-V
characteristics observed experimentally and correlate their features
to the specific properties of the molecule.

Our approach naturally accounts for the experimental observations and
indicates that the current in these molecular junctions is mainly
controlled by the electronic structure of the molecules and their
local environment. We demonstrate that many molecular orbitals
participate in a single conduction channel and provide examples where,
surprisingly, the current is not dominated by the contribution from
the energetically closest MOs. The theory provides a
quantitative criterion to judge the importance of individual MOs to
the current and thus paves the way for the {\em a priori} design of
molecular transport properties.

{\em Theoretical model.}--- We calculate the current through a single
molecule attached to metallic electrodes by a generalization of an
earlier analysis of transport in atomic-size contacts~\cite{Cuevas98},
similar in spirit to Refs.~\cite{Yaliraki99,Palacios01}. Since the
conductance is mainly determined by the narrowest part of the
junction, only the electronic structure of this ``central cluster'' must be
resolved in detail. It is therefore sensible to decompose the overall
Hamiltonian of the molecular junction as
\begin{equation}
\hat{H} = \hat{H}_L + \hat{H}_R + \hat{H}_C + \hat{V},
\end{equation}
where $\hat{H}_C$ describes the ``central cluster" of the system,
$\hat{H}_{L,R}$ describe the left and right electrode respectively, and
$\hat{V}$ gives the coupling between the electrodes and the central
cluster (see Fig.~1).

In this study the electronic structure of the ``central cluster'' is
calculated within the density functional (DFT) approximation~\cite{DFT}.
The left and right reservoirs are modeled as two perfect
semi-infinite crystals of the corresponding metal using a
tight-binding parameterization~\cite{Papa86}. Finally,  $\hat{V}$ 
describes the coupling between the leads and the central cluster and
takes the form: $\hat{V} = \sum_{ij} v_{ij} (\hat{d}^{\dagger}_i
\hat{c}_j + h.c.)$. The hopping elements $v_{ij}$ between
the lead orbitals $\hat{d}^{\dagger}_i$ and the MOs of the central
cluster $\hat{c}^{\dagger}_j$ are obtained by reexpressing
the MOs of the central cluster via a L\"owdin transformation in terms
of atom-like orbitals, and then using the mentioned tight-binding
parameterization~\cite{Papa86}.

The ``central cluster'' is not necessarily confined to the molecule,
but may, in principle, contain arbitrary parts of the metallic
electrode. The inclusion of part of the leads in the {\em ab initio}
calculation was shown to improve the description of the molecule-leads
coupling~\cite{Ventra00}, in particular regarding charge transfer
between the molecule and the electrodes. The Fermi energy of the overall
system is determined by the charge neutrality condition of the central
cluster.

\begin{figure}[t]
\begin{center}
\includegraphics[width=\columnwidth]{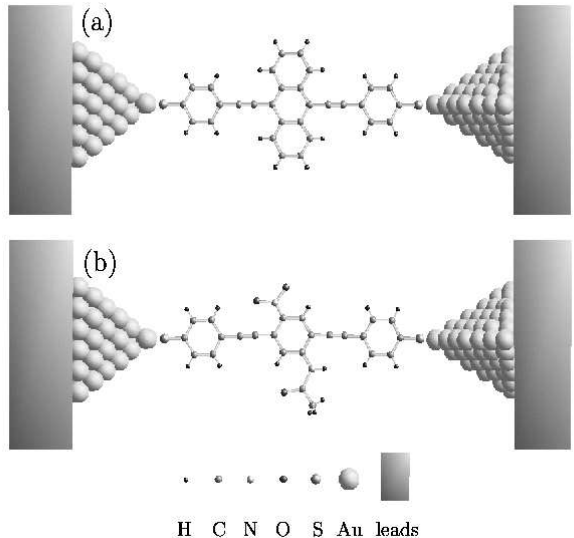}
\caption{\label{setup} Scheme of the 
single-molecule contacts analyzed in this work. The two organic molecules
attached to gold electrodes, which were experimentally investigated in 
Ref.~\cite{Reichert01}, are referred to as: (a) ``symmetric molecule", and (b) 
``asymmetric molecule". } 
\end{center}
\end{figure}

In order to obtain the current for a constant bias voltage, $V$,
between the leads, we make use of non-equilibrium Green function
techniques. Since the Hamiltonian of Eq.~(1) does not contain
inelastic interactions, the current follows from the Landauer
formula~\cite{Landauer}
\begin{equation}
\hspace*{-2.5mm} I = \frac{2e}{h} \int^{\infty}_{-\infty} d\epsilon \;
\mbox{Tr} \left\{ \hat{t} \hat{t}^{\dagger} \right\} 
\left[f(\epsilon-eV/2) - f(\epsilon+eV/2) \right],
\end{equation}

\noindent
where $f$ is the Fermi function and $\hat{t}$ is the energy and
voltage dependent transmission matrix given by

\begin{equation}
\hat{t}(\epsilon,V) = 2 \; \hat{\Gamma}^{1/2}_L(\epsilon-eV/2) 
\hat{G}^r_{C}(\epsilon,V) \hat{\Gamma}^{1/2}_R(\epsilon+eV/2) . 
\end{equation}

\noindent
The scattering rate matrices are given by $\hat{\Gamma}_{L,R} =
\mbox{Im} ( \hat{\Sigma}_{L,R} )$, where $\hat{\Sigma}_{L,R}$ are the
self-energies which contain the information of the electronic
structure of the leads and their coupling to the central cluster. They
can be expressed as $\hat{\Sigma}_{L,R} (\epsilon) = \hat{v}_{CL,R}
g_{L,R}(\epsilon) \hat{v}_{L,RC}$, $\hat{v}$ being the hopping
matrix which describes the connection between the central cluster and
the leads. $g_{L,R}$ are the Green functions of the uncoupled
leads (semi-infinite crystals), which are calculated using decimation
techniques~\cite{Guinea83}. The Green functions of the central
cluster are given by
\begin{equation}
\hat{G}_C(\epsilon,V) = \left[ \epsilon \hat{1} - \hat{H}_C - 
\hat{\Sigma}_L (\epsilon-eV/2) - \hat{\Sigma}_R (\epsilon+eV/2) \right]^{-1} .
\end{equation}
In order to provide a deep understanding of the electronic transport, 
we analyze the current in terms of {\em conduction channels}, defined
as eigenfunctions of $\hat{t} \hat{t}^{\dagger}$. Such analysis
allows to quantify the contribution to the transport of {\em every
individual molecular level}. In our approach the channels arise as a 
linear combination of the molecular orbitals $|\phi_j\rangle $ of 
the central cluster, i.e. $|c \rangle = \sum_j \alpha_{cj} |\phi_j\rangle$, 
and the corresponding eigenvalues determine their contribution to the
conductance. Ultimately, this information concerning the channels could 
eventually be measured using superconducting electrodes~\cite{Scheer98}.

\begin{figure}[!ht]
\begin{center}
\includegraphics[width=\columnwidth]{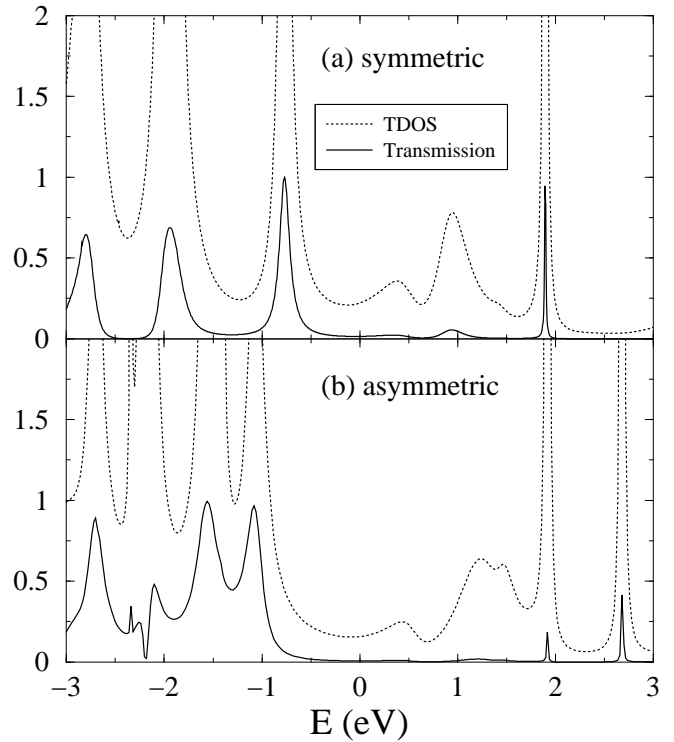}
\caption{\label{DOS} Total density of states (TDOS) of the molecule and 
zero-bias total transmission as a function of the energy for both molecules. 
The Fermi energy is set to zero.}
\end{center}
\end{figure}

\begin{figure*}[t]
\includegraphics*[width=\textwidth]{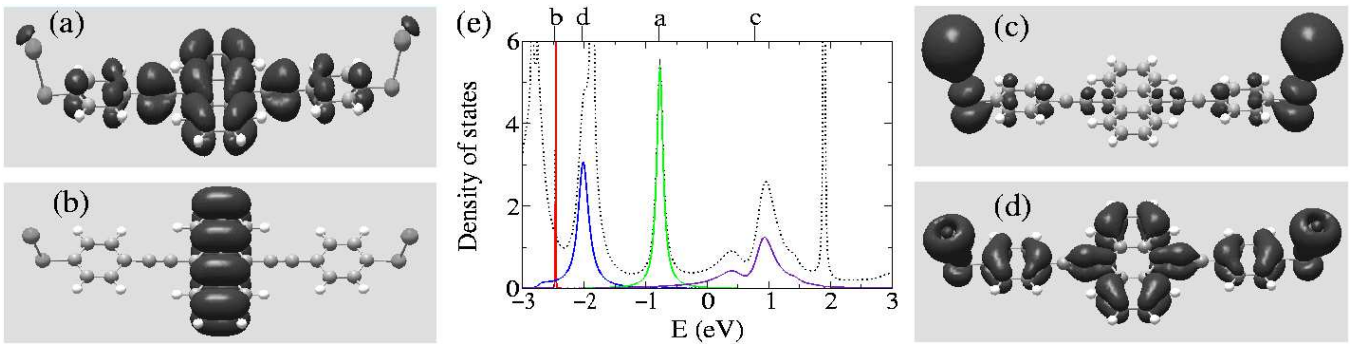}
\caption{\label{MO} (a-d) Charge-density plots of four molecular orbitals
of the central cluster for the symmetric molecule. Panel (a) 
displays the HOMO and (c) the LUMO, which is twofold degenerate. (b) shows 
a confined orbital that contributes little to the current, while the MO in 
(d) is almost as important as the LUMO despite its difference in energy. Panel 
(e) shows the total density of states of the central cluster (dotted line) and 
the individual contributions of the four molecular orbitals (color lines). 
The level positions are indicated on top of this panel. The contributions 
of the different MOs to the conduction channel at the Fermi energy (set to
zero) are: $|\alpha_a|^2 = 0.007$, $|\alpha_b|^2 = 10^{-11}$, $|\alpha_c|^2 = 
0.06$, $|\alpha_d|^2 = 0.02$.}
\end{figure*}

\begin{figure}[!b]
\begin{center}
\includegraphics[width=0.8\columnwidth]{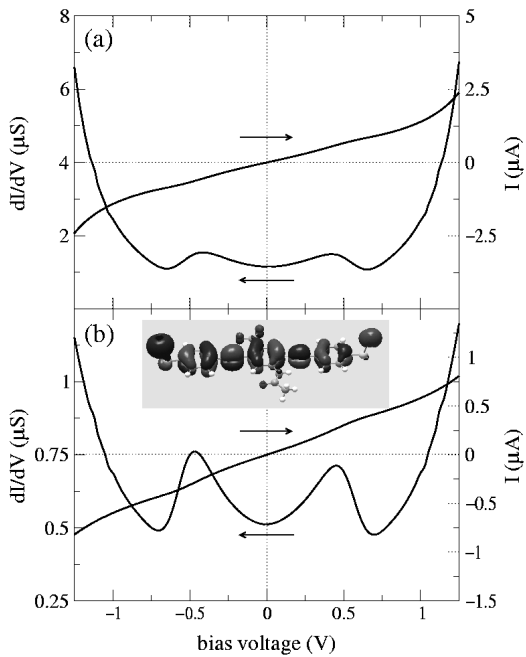}
\caption{\label{I-V} I-V characteristics and differential conductance for the 
symmetric (a) and asymmetric (b) molecules. The inset in (b) shows the 
charge-density plot of the HOMO for the asymmetric molecule. 
Notice the intrinsic asymmetry of the charge distribution in the gold atoms.}
\end{center}
\end{figure}

{\em Results and discussions.}--- We now use the method described
above to analyze the experiment of Reichert \emph{et al.}~\cite{Reichert01}. 
The two organic molecules investigated are shown in
Fig.~1, and as indicated in the caption, they will be referred to as
symmetric and asymmetric molecule.  For the description of the gold
reservoirs, we use a basis with the atomic-like $5d,6d,6p$ orbitals,
and for the central cluster we use the LANL2DZ basis~\cite{LANL2DZ}
for all atoms. The DFT calculations were performed with the Becke
three-parameter hybrid functional using the Lee, Yang and Parr
correlation functional~\cite{Becke93} at zero field~\cite{fieldcomment}. 
In the calculations reported here one additional gold atom was included on 
either side of the molecule. Experimentally, both molecules were contacted 
several times and the nature of the I-V characteristics was found to vary with 
the quality of the contact. For this reason, theory can presently aim to 
elucidate important reproducible features of the experiment under the 
assumption that the contact to the electrodes is well defined. Since there 
is no direct experimental information regarding the geometry of the molecule 
and its attachment to the leads, the overall geometry of the central cluster 
was relaxed without additional constraints in our calculations, resulting in 
the Au atom being out of the molecular plane.

Let us start by analyzing the linear response regime. In Fig.~2 we
show for both cases the total density of states (TDOS) of the molecule
and the zero-bias total transmission as a function of energy. As can
be seen in the TDOS, in both cases the covalent bond between Au and S
results in a strong hybridization between the molecular orbitals and
the extended states of the metallic electrodes.  The formation of wide
energy bands and the disappearance of the gaps of the discrete
molecular spectrum suggest the absence of Coulomb blockade in this
type of molecular junctions.

The zero-bias total transmission as a function of energy follows
closely the TDOS.  The transmission is dominated overwhelmingly by a
single channel in the energy window shown in Fig.~2, and the
corresponding eigenvalues of $\hat{t} \hat{t}^{\dagger}$ at the Fermi
energy~\cite{fermiref} are $T_{sym} = 0.014$ and $T_{asym} = 0.006$. 
The decomposition of this channel into molecular orbitals provides us 
information on the relevance of the different molecular levels. Fig.~3 
(a)-(d) show charge-density plots for some representative MOs and (e) shows 
their individual contribution to the TDOS. Fig.~3a shows that the 
highest occupied molecular orbital (HOMO) is confined to the interior 
of the molecule and its weight at the gold atoms is rather small. 
Consequently, in spite of its privileged energy position, the HOMO 
does not give a significant contribution to the current. The lowest 
unoccupied molecular orbital (LUMO), see Fig.~3c, exhibits the opposite 
behavior, i.e. it is very well coupled to the leads through the $6s$ atomic 
orbital of the gold atoms (notice that it has width of about 4 $eV$ in the 
density of states), but the charge is mainly localized on the Au and S atoms. 
The interplay of these two factors yields a contribution of $\approx 6\%$ of 
the total current. Fig.~3(b,d) shows two further MOs with similar energy 
but very different contribution to the channel. While the localized MO (b) 
carries almost no current, the extended and well coupled MO (d) has
significant weight. Consequently there are three ingredients which
determine the contribution of a MO to the current: (i) its energy
position (distance to the Fermi energy), (ii) its bridging extent
(whether it is extended or localized), and (iii) its coupling to the
leads. Our analysis provides a counterexample to the conventional
wisdom that the HOMO and the LUMO dominate the transport properties.

Fig.~4 shows the I-V curves for both molecules in the voltage range
investigated in Ref.~\cite{Reichert01}. Both the order of
magnitude and shape of the current and conductance agree qualitatively
with the experimental results.  There is no pronounced voltage
dependence of the transmission due to the smooth density of states
of the gold electrodes in the energy region explored here.  The
non-linearities in these I-V curves can be then understood by a simple
inspection of the energy dependence of the zero-bias transmission.
For instance, the pronounced increase in the conductance around 1 V is
due to the fact that we approach the resonant condition for the
HOMO and LUMO.

In agreement with the experiment, the I-V of the symmetric molecule is
symmetric with respect to voltage inversion, while the one of the
asymmetric molecule is asymmetric. According to Eq.~2 we note that the
asymmetry of the MO is not a sufficient condition for an asymmetric I-V
characteristic, as evidenced by the experimental I-V near zero bias.
Only the energy dependence of the left/right scattering rates in Eq.~3
induces an asymmetry into the I-V characteristic. Such asymmetries can
arise either from differences in the local atomic structure of the electrodes 
or from intrinsic properties of the molecule. The latter arise from the
asymmetric charge distribution of the MOs (see inset Fig.~4b), resulting 
in different couplings to the leads. Even for the symmetric
molecule, we were able to induce asymmetries into the I-V characteristic 
by distorting the geometry of one of the lead fragments in the central 
cluster. This fact was nicely demonstrated in the experiment
(see Fig.~5 in Ref.~\cite{Reichert01}). We investigated
several other scenarios regarding the number of gold atoms, their
geometry and the coupling and found predictable variations of the I-V's
with theses changes. If, for example, the gold-sulfur bond is
stretched by about 5\%, the current changes by about 70\%, but the
position of the peaks in the conductance and their peak/valley ratios
are only marginally affected.

{\em Conclusions.}--- We have presented an atomistic semi-quantitative
description of non-linear transport through a single molecule
junction. We were able to attribute distinctive features of the I-V's
of the symmetric and asymmetric molecule to their individual molecular
levels obtained from {\em ab initio} calculations. The resolution of
conductance into conduction channels permits an analysis of
the contributions of individual orbitals to overall transport. 
In this way we provide an understanding which can be valuable
for the future engineering of molecular devices.

We are grateful for many stimulating discussions with D. Beckmann, M. Hettler,
M. Mayor, H. Weber and F. Weigend. The work is part of the CFN which is 
supported by the DFG. JCC acknowledges funding by the EU TMR Network on 
Dynamics of Nanostructures, and WW by the German National Science Foundation 
(We 1863/10-1), the BMBF and the von Neumann Institute for Computing. 



\begin{thebibliography}{}
\bibitem{fermiref} We chose the Fermi energy in the middle of the
HOMO/LUMO gap and verified that for the molecules investigated here 
the transmission at the Fermi energy is not very sensitive to this choice.
\vspace*{-8mm}
\end{thebibliography}
\end{document}